\documentstyle[12pt,epsfig]{article}

\catcode`\@=11
\def\@citex[#1]#2{%
\if@filesw \immediate \write \@auxout {\string \citation {#2}}\fi
\@tempcntb\m@ne \let\@h@ld\relax \def\@citea{}%
\@cite{%
  \@for \@citeb:=#2\do {%
    \@ifundefined {b@\@citeb}%
      {\@h@ld\@citea\@tempcntb\m@ne{\bf ?}%
      \@warning {Citation `\@citeb ' on page \thepage \space undefined}}%
      {\@tempcnta\@tempcntb \advance\@tempcnta\@ne%
      \@tempcntb\number\csname b@\@citeb \endcsname \relax%
      \ifnum\@tempcnta=\@tempcntb 
        \ifx\@h@ld\relax%
          \edef \@h@ld{\@citea\csname b@\@citeb\endcsname}%
        \else%
          \edef\@h@ld{\ifmmode{-}\else--\fi\csname b@\@citeb\endcsname}%
        \fi%
      \else
        \@h@ld\@citea\csname b@\@citeb \endcsname%
        \let\@h@ld\relax%
      \fi}%
    \def\@citea{,\penalty\@highpenalty\,}%
  }\@h@ld
}{#1}}

\def\@citeb#1#2{{[#1]\if@tempswa , #2\fi}}
\def\@citeu#1#2{{$^{#1}$\if@tempswa , #2\fi }}
\def\@citep#1#2{{#1\if@tempswa , #2\fi}}

\def\bcites{         
        \catcode`\@=11
        \let\@cite=\@citeb
        \catcode`\@=12
}

\def\upcites{         
        \catcode`\@=11
        \let\@cite=\@citeu
        \catcode`\@=12
}

\def\plaincites{      
        \catcode`\@=11
        \let\@cite=\@citep
        \catcode`\@=12
}

\newcount\hour
\newcount\minute
\newtoks\amorpm
\hour=\time\divide\hour by 60
\minute=\time{\multiply\hour by 60 \global\advance\minute by-\hour}
\edef\standardtime{{\ifnum\hour<12 \global\amorpm={am}%
        \else\global\amorpm={pm}\advance\hour by-12 \fi
        \ifnum\hour=0 \hour=12 \fi
        \number\hour:\ifnum\minute<10 0\fi\number\minute\the\amorpm}}
\edef\militarytime{\number\hour:\ifnum\minute<10 0\fi\number\minute}

\def\draftlabel#1{{\@bsphack\if@filesw {\let\thepage\relax
   \xdef\@gtempa{\write\@auxout{\string
      \newlabel{#1}{{\@currentlabel}{\thepage}}}}}\@gtempa
   \if@nobreak \ifvmode\nobreak\fi\fi\fi\@esphack}
        \gdef\@eqnlabel{#1}}
\def\@eqnlabel{}
\def\@vacuum{}
\def\marginnote#1{}
\def\draftmarginnote#1{\marginpar{\raggedright\scriptsize\tt#1}}
\def\draft{
        \pagestyle{plain}
        \overfullrule=2pt
        \oddsidemargin -.5truein
        \def\@oddhead{\sl \phantom{\today\quad\militarytime} \hfil
        \smash{\Large\sl DRAFT} \hfil \today\quad\militarytime}
        \let\@evenhead\@oddhead
        \let\label=\draftlabel
        \let\marginnote=\draftmarginnote
        \def\ps@empty{\let\@mkboth\@gobbletwo
        \def\@oddfoot{\hfil \smash{\Large\sl DRAFT} \hfil}
        \let\@evenfoot\@oddhead}
        \def\@eqnnum{(\theequation)\rlap{\kern\marginparsep\tt\@eqnlabel}%
        \global\let\@eqnlabel\@vacuum}  }

\def\eqalign#1{\null\,\vcenter{\openup\jot\m@th
  \ialign{\strut\hfil$\displaystyle{##}$&$\displaystyle{{}##}$\hfil
      \crcr#1\crcr}}\,}
\def\eqalignno#1{\displ@y \tabskip\centering
  \halign to\displaywidth{\hfil$\@lign\displaystyle{##}$\tabskip\z@skip
    &$\@lign\displaystyle{{}##}$\hfil\tabskip\centering
    &\llap{$\@lign##$}\tabskip\z@skip\crcr
    #1\crcr}}

\def\section{\@startsection {section}{1}{\z@}{3.ex plus 1ex minus
 .2ex}{2.ex plus .2ex}{\large\bf}}
\def\subsection{\@startsection{subsection}{2}{\z@}{2.75ex plus 1ex minus
 .2ex}{1.5ex plus .2ex}{\bf}}

\def\appendix#1{
  \addtocounter{section}{1}
  \setcounter{equation}{0}
  \renewcommand{\thesection}{\Alph{section}}
  \section*{Appendix \thesection\protect\indent \parbox[t]{11.15cm}
  {#1} }
  \addcontentsline{toc}{section}{Appendix \thesection\ \ \ #1}
  }

\def\abstract{\if@twocolumn
\section*{Abstract}
\else 
\begin{center}
{\bf Abstract\vspace{-.5em}\vspace{0pt}}
\end{center}
\quotation
\fi}

\catcode`\@=12

\catcode`\@=11
\def\theequation{\arabic{equation}}
\catcode`\@=12

\bcites

\catcode`\@=11
\def\theequation{\thesection.\arabic{equation}}
\@addtoreset{equation}{section}
\catcode`\@=12

\newcommand{\beq}{\begin{equation}}
\newcommand{\beqa}{\begin{eqnarray}}
\newcommand{\bega}{\begin{array}}
\newcommand{\ea}{\end{array}}

\newcommand{\eeq}{\end{equation}}
\newcommand{\eeqa}{\end{eqnarray}}

\newcommand{\n}{\nabla}

\newcommand{\lra}{{\leftrightarrow}}

\newcommand{\NN}{\mbox{$\cal N$}}

\newcommand{\bone}{\mbox{\bf 1}}
\newcommand{\tr}{\mbox{tr$\,$}}

\newcommand{\half}{\mbox{$1\over2$}}

\newcommand{\OO}{\mbox{$\cal O$}}

\newcommand{\RR}{\mbox{$\cal R$}}

\newcommand{\LL}{\mbox{$\cal L$}}

\newcommand{\e}{\epsilon}

\newcommand{\sch}{Schwarzschild}

\begin{document}

\vskip-1pt
\hfill {\tt gr-qc/0505039}
\vskip-1pt
\hfill BRX TH-563
\vskip0.2truecm
\begin{center}
\vskip 0.2truecm {\Large\bf
Curvature invariants of 
\\ \vskip 0.2truecm
static spherically symmetric geometries
}
\vskip 1.2truecm
{\bf
S. Deser\footnote{Email: deser@brandeis.edu}
and 
A.V. Ryzhov\footnote{E-mail: ryzhovav@brandeis.edu}\\
\vskip 0.4truecm
{\it  Department of Physics,  Brandeis University\\
Waltham, MA 02454, USA}\\
}
\end{center}
\vskip 0.5truecm
\vskip 0.2truecm \noindent\centerline{\bf Abstract}
\vskip .2truecm

We construct all independent local scalar monomials 
in the Riemann tensor at arbitrary dimension,
for the special regime of static, spherically symmetric geometries. 
Compared to general spaces, their number is significantly reduced:
the extreme example is the collapse of all invariants $\sim$ Weyl${}^k$, 
to a single term at each $k$.
The latter is equivalent to the Lovelock invariant $\LL_k$.
Depopulation is less extreme for invariants involving rising numbers of Ricci tensors, 
and also depends on the dimension.
The corresponding local gravitational actions and their solution spaces
are discussed.

\newpage
\setcounter{footnote}{0}

\section{Introduction}
\label{section: intro}

Our subject, construction and enumeration of local curvature invariants for special geometries, 
i.e. scalars of the form $(R_{\mu\nu\alpha\beta})^k$ in arbitrary dimension $D$,
has both mathematical and physical interest.
Two physical applications of the general scheme stand out:
the first is to understand geometrical corrections to general relativity, 
which is ``merely'' the lowest order term of an effective series 
of loop or string corrections.
For this purpose, an accessible grid of candidate terms is desirable, 
and indeed an enormous literature has accumulated 
just on the three quadratic possibilities, 
$R^2$, $R_{\mu\nu} R^{\mu\nu}$, $R_{\mu\nu\alpha\beta} R^{\mu\nu\alpha\beta}$.
A second, related, application is to list allowed 
local counterterms in specific quantum models, such as supergravity.
The first example, that of $D=4$, $\NN=1$ SUGRA, produced an allowed supersymmetric 
counterterm, 
the square of the quadratic Bel-Robinson tensor 
$T_{\mu\nu\alpha\beta} \sim (R R)_{\mu\nu\alpha\beta}$
at 3-loop order \cite{Deser:1977nt}.
More recently, it became possible to actually compute explicitly 
the 2-loop divergences in $D=11$ SUGRA \cite{Bern:1998ug}, the bosonic part of which 
was also quartic in curvature (and in four-form field strength).
Here again, knowledge of the basis for quartics was a useful tool \cite{Deser:1999me,Deser:2000xz}.
Meanwhile, on the mathematical side, 
an algorithmic way to find all independent monomial scalars in Riemann${}^k$,
along with a number of illuminating explicit examples,%
\footnote{
Unlike \cite{Fulling92}, 
we will not consider invariants involving explicit derivatives, 
such as $R^n \nabla^{2p} R^m$, although they will certainly appear in any 
physical context alongside their algebraic counterparts $R^{n+m+p}$. 
} 
was given in \cite{Fulling92}. 
That work illustrated the discouragingly rapid rise of their number with $k$, 
and to a lesser extent, $D$.

Once the basis of possible corrections is established,
there are a number of immediate physical questions one may ask. 
The most obvious is that of the solutions to the effective actions:
will they provide surprises about horizons, naked singularities and other post-\sch ean behavior?
Here, of course, the only hope of obtaining solutions 
lies in the simplest possible geometries,
namely the static spherically symmetric ones;
this question is one of our major motivations.
So, apart from a brief excursion away from 
static spherically symmetric geometries 
in Section \ref{section: cylindrical}
to show how intractable things become for less symmetric spaces,
all work below is restricted to static spherically symmetric geometries.

We will establish a decomposition of the curvature tensor 
into the Weyl and Ricci parts 
to provide a simple analysis of 
all generic local curvature invariants.
As a result, we will note a dramatic decrease 
in the number of independent terms 
and an increase in their transparency.
In particular, the general classification will be 
governed more by the relative powers of Weyl and Ricci components,
rather than by the overall monomial power. 
We will then be able to discuss classes, 
rather than individual terms, 
thereby helping navigate the sea of effective actions and their solutions.

\section{Symmetries and definitions}
\label{section: ansatz}

We first introduce notation that will make the basis elements 
$(R_{\mu\nu\alpha\beta})^k$
as easily computable as possible in arbitrary dimensions%
\footnote{
In $D=3$, the Weyl tensor vanishes identically, 
Riemann and Ricci being equivalent (double duals) to each other;
in $D=2$, only the scalar curvature $R_\mu{}^\mu$ remains. 
} 
$D=n+2 \ge 4$.
Our metric is expressed in \sch{} gauge:
\beqa
\label{eq: metric ansatz}
ds^2 
= - e^{2 \Phi(r)} dt^2 + e^{2 \Lambda(r)} dr^2 + r^2 d\Omega^2_n
= - \omega_{\hat t}^2 + \omega_{\hat r}^2 + \sum_{i=1}^n \omega_{\hat i} \omega_{\hat i} 
\eeqa
where the orthonormal basis one-forms are
\beqa
\label{eq: orthonormal forms}
\omega_{\hat t} = -e^{\Phi} dt
,\quad 
\omega_{\hat r} = e^{\Lambda} dr
.
\eeqa
%
The curvature 2-form is
\beqa
\label{eq: curvature 2 form}
\RR^\mu{}_\nu 
\equiv d \omega^\mu{}_\nu +  \omega^\mu{}_\alpha \wedge \omega^\alpha{}_\nu 
= {1\over 2} R^\mu{}_{\nu\alpha\beta} \omega^\alpha \wedge \omega^\beta 
,\quad
\omega^\mu{}_\nu = \Gamma^\mu{}_{\nu\alpha} \omega^\alpha 
,\quad
\omega_{\mu\nu} = g_{\mu\alpha} \omega^\alpha{}_\nu 
.
\eeqa
Using the identities
\beqa
\label{eq: omega id}
0 = d \omega^\mu + \omega^\mu{}_\nu \omega^\nu
,\quad
d g_{\mu\nu} = \omega_{\mu\nu} + \omega_{\nu\mu}
\eeqa
and the orthonormal basis (\ref{eq: orthonormal forms}), 
it is easy to find the connection one-forms $\omega_{\hat \mu \hat \nu} = - \omega_{\hat \nu \hat \mu}$:
\beqa
\label{eq: omega solved}
\omega_{\hat t \hat r} = - \Phi' e^{-\Lambda} \omega^{\hat t}
,\quad
\omega_{\hat i \hat r} = {1\over r} e^{-\Lambda} \omega^{\hat i}
,\quad
\omega_{\hat t \hat i} = 0
.
\eeqa
The resulting $\RR^\mu{}_\nu$ for $D=4$ are given in \cite{MTW}:
\beqa
\label{eq: curvature 2form solved}
\RR^{\hat t}{}_{\hat r} = 
e^{-2\Lambda} \left[ \Phi'' + (\Phi')^2 - \Phi' \Lambda' \right] \omega^{\hat r} \wedge \omega^{\hat t}
&\equiv&
A(r) \omega^{\hat r} \wedge \omega^{\hat t}
,\nonumber\\
\RR^{\hat t}{}_{\hat i} = 
e^{-2\Lambda} \left[ - {1\over r} \Phi' \right] \omega^{\hat t} \wedge \omega^{\hat i}
&\equiv&
B(r) \omega^{\hat t} \wedge \omega^{\hat i} 
,\nonumber\\
\RR^{\hat r}{}_{\hat i} =
e^{-2\Lambda} \left[ {1\over r} \Lambda' \right] \omega^{\hat r} \wedge \omega^{\hat i}
&\equiv&
C(r) \omega^{\hat r} \wedge \omega^{\hat i} 
,\nonumber\\
\RR^{\hat i}{}_{\hat j} = 
{1\over r^2} \left[ 1 - e^{-2\Lambda} \right] \omega^{\hat i} \wedge \omega^{\hat j} 
&\equiv&
\psi(r) \omega^{\hat i} \wedge \omega^{\hat j}
;
\eeqa
they are in fact dimension-independent.
The functions $A$, $B$, $C$, and $\psi$, 
defined in (\ref{eq: curvature 2form solved}), 
are the only combinations of $\Lambda(r)$ and $\Phi(r)$
appearing in the Riemann tensor.

The $\{ A, B, C, \psi \}$ are algebraically, 
but not functionally, independent; 
for example, 
$C(r) = {1\over 2 r} (r^2 \psi(r))'$.
When $A$ and $B$ are used for constructing actions, 
we can integrate them by parts:
\beqa
\label{eq: A and B by parts}
\int_{0}^{\infty} dr r^n e^{\Phi+\Lambda} B \Psi
&=&
\int_{0}^{\infty} dr r^n e^{\Phi+\Lambda} 
\left\{ 
\left( {1\over r^2} - \psi \right) \left[ (n-1) \Psi + r \Psi' \right] - C \Psi
\right\}
,
\nonumber\\
\int_{0}^{\infty} dr r^n e^{\Phi+\Lambda} A \Psi
&=&
\int_{0}^{\infty} dr r^n e^{\Phi+\Lambda} 
\left\{ 
B \left[ n \Psi + r \Psi' \right] 
\right\} 
\eeqa
where $\Psi=\Psi(r)$ is an arbitrary function.
The following operator relations allow us to express 
any monomial involving either a single $A$ or a single $B$
(but not both, and no higher powers of these) 
and an arbitrary number of $C$'s and $D$'s 
in terms of the single function $\psi$ and its derivatives:
\beqa
\label{eq: A B C of psi}
C = \psi + {1\over 2} r \psi'
,~
B \sim
\left( {1\over r^2} - \psi \right) \left[ (n-1) + r {d \over dr} \right] - C
,~
A \sim
B \left[ n + r {d \over dr} \right] 
\eeqa
where ``$\sim$'' means ``up to integrating by parts with measure $dr r^n e^{\Phi+\Lambda}$.''
The Einstein action is immediately obtained using (\ref{eq: A B C of psi}), 
\beqa
\label{eq: Einstein action}
S_1 = \int \sqrt{g} (R + \lambda)
=
\int dt \int d^n \Omega \int_{0}^{\infty} dr e^{\Phi+\Lambda} 
\left[ \left(n \psi(r) + {1\over n+1} \lambda \right) r^{n+1} \right]'
\eeqa
whose extrema%
\footnote{
This reduction, justified in \cite{Palais}, 
has also proved quite useful for quadratic curvature models \cite{Deser:2003up}.
}
are obviously the \sch-deSitter metrics 
\beqa
\label{eq: Einstein solution}
\Phi = - \Lambda = 
{1\over 2} \ln \left[ 1 - {\zeta \over r^{n-1}} + {\lambda r^2 \over n (n+1)} \right]
.
\eeqa

\section{Projectors and components}
\label{section: projectors}

The metric (\ref{eq: metric ansatz}) separates the space-time indices $\mu$
into $t$, $r$, and the angular $i=1, ..., n$.
The tensors 
\beqa
\label{eq: projectors}
\tau_{\hat \mu}{}^{\hat \nu} &\equiv& \delta_{\hat \mu}^{\hat t} \delta_{\hat t}^{\hat \nu}
= {\rm diag}(1,0,0,...,0)
,\nonumber\\
\rho_{\hat \mu}{}^{\hat \nu} &\equiv& \delta_{\hat \mu}^{\hat r} \delta_{\hat r}^{\hat \nu}    
= {\rm diag}(0,1,0,...,0)
,\nonumber\\
\sigma_{\hat \mu}{}^{\hat \nu} &\equiv& \sum_{i=1}^n \delta_{\hat \mu}^{\hat i} \delta_{\hat i}^{\hat \nu}
= {\rm diag}(0,0,1,...,1)
\eeqa
conveniently summarize this decomposition.
They are orthogonal projectors,
\beqa
\label{eq: projectors: orthonormality}
\tau \tau = \tau
,\quad
\rho \rho = \rho 
,\quad
\sigma \sigma = \sigma 
;\quad
\tau \rho = \tau \sigma = \rho \sigma = 0
,
\eeqa
and multiply as matrices, 
$(\tau \tau)_{\hat \mu}{}^{\hat \alpha} = 
\tau_{\hat \mu}{}^{\hat \nu} \tau_{\hat \nu}{}^{\hat \alpha} = 
\tau_{\hat \mu}{}^{\hat \alpha}$ 
etc.
Their traces are 
\beqa
\label{eq: projectors: traces}
\tr \tau = 1
,\quad
\tr \rho = 1 
,\quad
\tr \sigma = n 
.
\eeqa
In an orthonormal frame, indices 
are raised and lowered by the Minkowski metric, 
e.g. $\tau_{\hat \mu \hat \nu} = \tau_{\hat \mu}{}^{\hat \alpha} \eta_{\hat \alpha \hat \nu}$.
Since the (\ref{eq: projectors}) are symmetric matrices, 
we need not worry about index order, and can write
$\tau_{\hat \mu}^{\hat \nu}$ for $\tau_{\hat \mu}{}^{\hat \nu}$, etc.

The projectors (\ref{eq: projectors}) also 
provide a compact notation for the curvatures.
The Riemann tensor is
\beqa
\label{eq: Riemann: projectors}
R^{\hat \mu \hat \nu}{}_{\hat \alpha \hat \beta} &=& 
2 \left[
- A \cdot 2 \tau_{[\hat \alpha}^{[\hat \mu} \rho_{\hat \beta]}^{\hat \nu]} 
+ B \cdot 2 \tau_{[\hat \alpha}^{[\hat \mu} \sigma_{\hat \beta]}^{\hat \nu]} 
+ C \cdot 2 \rho_{[\hat \alpha}^{[\hat \mu} \sigma_{\hat \beta]}^{\hat \nu]} 
+ \psi \cdot \sigma_{[\hat \alpha}^{[\hat \mu} \sigma_{\hat \beta]}^{\hat \nu]} 
\right]
, 
\eeqa
while Ricci and Weyl can be written as
\beqa
\label{eq: Ricci: projectors}
R_{\hat \mu}{}^{\hat \nu} 
&=& 
F \tau_{\hat \mu}^{\hat \nu} + G \rho_{\hat \mu}^{\hat \nu} + H \sigma_{\hat \mu}^{\hat \nu}
,
\\
\label{eq: Weyl: projectors}
C^{\hat \mu \hat \nu}{}_{\hat \alpha \hat \beta} &=& 
-2 \chi {n-1 \over n+1} 
\left[ 
2 \tau_{[\hat \alpha}^{[\hat \mu} \rho_{\hat \beta]}^{\hat \nu]} 
- 2 {1\over n} \left( \tau_{[\hat \alpha}^{[\hat \mu} \sigma_{\hat \beta]}^{\hat \nu]} 
+ \rho_{[\hat \alpha}^{[\hat \mu} \sigma_{\hat \beta]}^{\hat \nu]} \right) 
+ {2\over n(n-1)} \sigma_{[\hat \alpha}^{[\hat \mu} \sigma_{\hat \beta]}^{\hat \nu]} 
\right]
\eeqa
where 
\beqa
\label{eq: ricci component functions}
F \equiv n B - A,~ G \equiv n C - A,~ H \equiv B + C + (n-1) \psi, 
\eeqa
are the Ricci components, and 
\beqa
\label{eq: weyl component function}
\chi \equiv A + B + C - \psi
\eeqa
is the only function appearing in the definition of Weyl.
Here, 
the brackets denote (normalized) antisymmetrization, 
$2 \sigma_{[\hat \alpha}^{[\hat \mu} \sigma_{\hat \beta]}^{\hat \nu]} =
\sigma_{\hat \alpha}^{\hat \mu} \sigma_{\hat \beta}^{\hat \nu} - 
\sigma_{\hat \beta}^{\hat \mu} \sigma_{\hat \alpha}^{\hat \nu}$, 
etc.
The (four-index) structures 
$2 \tau_{[\hat \alpha}^{[\hat \mu} \rho_{\hat \beta]}^{\hat \nu]}$, 
$2 \tau_{[\hat \alpha}^{[\hat \mu} \sigma_{\hat \beta]}^{\hat \nu]}$, 
$2 \rho_{[\hat \alpha}^{[\hat \mu} \sigma_{\hat \beta]}^{\hat \nu]}$, and  
$\sigma_{[\hat \alpha}^{[\hat \mu} \sigma_{\hat \beta]}^{\hat \nu]}$
are also orthogonal projectors,
e.g., 
$
\sigma_{[\hat \alpha}^{[\hat \mu} \sigma_{\hat \beta]}^{\hat \nu]}
\sigma_{[\hat \e}^{[\hat \alpha} \sigma_{\hat \gamma]}^{\hat \beta]}
=
\sigma_{[\hat \e}^{[\hat \mu} \sigma_{\hat \gamma]}^{\hat \nu]}
$. 
Powers of Riemann, Ricci, and Weyl have similar form
and are discussed in the Appendix. 

To list all independent invariants, 
we first organize them according to the number of Weyls (i.e., the power of $\chi$),
and then eliminate redundancies at every given power of $\chi$. 
The locations of the indices for Weyl and Ricci (before contractions into scalars)
can be fixed once and for all as 
$C^{\hat \mu \hat \nu}{}_{\hat \alpha \hat \beta}$ and $R_{\hat \mu}{}^{\hat \nu}$;
then, one can use (\ref{eq: Ricci: projectors}) and (\ref{eq: Weyl: projectors}) 
without having to worry about raising or lowering their indices.
All other configurations are linear combinations of these. 
The resulting invariants may still be linearly dependent;
for example, we have not made use of the cyclic symmetry 
$R_{\mu[\nu\alpha\beta]} =0$. 
So one may need to eliminate the extras by hand.
Invariants made out of the Riemann tensor, or equivalently 
its Weyl and Ricci components, can be written in terms of products of (\ref{eq: projectors}),
and contracted using (\ref{eq: projectors: orthonormality})
and (\ref{eq: projectors: traces});
these are the only tools required for our calculations.
As (\ref{eq: Weyl: projectors}) is symmetric under $\tau \lra \rho$, 
the resulting invariants will be symmetric under $F \lra G$; 
see (\ref{eq: Ricci: projectors})-(\ref{eq: Weyl: projectors}).

\section{Lovelock invariants}
\label{section: general lovelock}

To connect with earlier studies of specific actions, 
it is convenient to formulate Weyl${}^k$ monomials 
in terms of the Euler-Gauss-Bonnet-Lovelock invariants 
\beqa
\label{eq: general lovelock}
\LL_k \equiv 
\delta_{\alpha_1 \beta_1 \alpha_2 \beta_2 ... \alpha_k \beta_k}^{\mu_1 \nu_1 \mu_2 \nu_2 ... \mu_k \nu_k}
R_{\mu_1 \nu_1}{}^{\alpha_1 \beta_1}
R_{\mu_2 \nu_2}{}^{\alpha_2 \beta_2}
...
R_{\mu_k \nu_k}{}^{\alpha_k \beta_k}
,
\eeqa
with 
\beqa
\label{eq: antisym tensor delta}
\delta_{\alpha_1 \alpha_2 ... \alpha_r}^{\mu_1 \mu_2 ... \mu_r} 
\equiv 
{1 \over r!} \sum_{\gamma \in S_r}
{\rm sign}(\gamma) 
\delta_{\alpha_{\gamma(1)}}^{\mu_1} \delta_{\alpha_{\gamma(2)}}^{\mu_2} ... \delta_{\alpha_{\gamma(r)}}^{\mu_r} 
,
\eeqa
see e.g. \cite{Banados:1993ur}. 
Here, $\delta$
is totally antisymmetric in its upper and lower indices separately,
and the normalization is chosen to make $\delta$ a projector:
$
\delta_{\mu_1 \mu_2 ... \mu_r}^{\nu_1 \nu_2 ... \nu_r} 
\delta_{\alpha_1 \alpha_2 ... \alpha_r}^{\mu_1 \mu_2 ... \mu_r} 
= 
\delta_{\alpha_1 \alpha_2 ... \alpha_r}^{\nu_1 \nu_2 ... \nu_r}$.
When $D = 2 k$, $\LL_k$ is a total divergence, 
and it vanishes identically for $D < 2 k$.

The $\LL_k$ are easily evaluated.
Antisymmetry of the Riemann tensor (\ref{eq: Riemann: projectors}) 
in its upper and lower indices, reduces (\ref{eq: general lovelock}) to sums of 
\beqa
\label{eq: antisym projectors}
\delta_{\beta_1 ... \gamma_1 ... \epsilon_1 ...}^{\mu_1 ... \nu_1 ... \alpha_1 ...} 
(\tau^{\beta_1}_{\mu_1} ... \tau^{\beta_p}_{\mu_p}) 
(\rho^{\gamma_1}_{\nu_1} ... \rho^{\gamma_q}_{\nu_q}) 
(\sigma^{\epsilon_1}_{\alpha_1} ... \sigma^{\epsilon_r}_{\alpha_r}) 
= 
{n! \over (p+q+r)! (1 - p)! (1 - q)! (n - r)!} 
.
\eeqa
Permutations mixing indices from different groups ($\{\mu \}$, $\{ \nu \}$, $\{ \alpha \}$) 
vanish, since the projectors $\tau$, $\rho$, and $\sigma$ are mutually orthogonal; 
their traces are given in (\ref{eq: projectors: traces}). 
For (\ref{eq: antisym projectors}) to be nonzero, we must have $2k = p+q+r \le n+2 = D$. 
From the explicit form of (\ref{eq: Riemann: projectors}) we find that $\LL_k$ 
is the specific combination 
\beq
\label{eq: lovelock: evaluated}
\LL_k 
=
\zeta_k^n
\left[
(n-2k+1) \psi^k 
+ 2 k (B+C) \psi^{k-1}
+ {2 k  \over (n-2k+2)} (2 (k-1) B C -A \psi) \psi^{k-2}
\right]
,
\eeq
where the prefactor 
$\zeta_k^n \equiv {2^k n! \over (2 k)! (n-2k+1)!}$ (so $\zeta_k^n = 0$ for $n+2 \le 2 k$).
Using (\ref{eq: A B C of psi}) to integrate by parts, 
the $\LL_k$ give rise to uniformly nice actions 
\beqa
\label{eq: nice action}
S_{\cal L} = \int d^D x \sqrt{g} \sum_k a_k \LL_k 
= \int dt \int d^n \Omega 
\int_{0}^{\infty} dr e^{\Phi+\Lambda} \left[ r^{n+1} \sum_k a_k \zeta_k^n \psi(r)^k \right]'
\eeqa
for any set of coefficients $a_k$. 
The attractive properties of the Einstein action (\ref{eq: Einstein action})
are preserved: ``on-shell,'' $\Phi + \Lambda$ is constant 
and $\psi(r)$ satisfies an algebraic (rather than a differential) equation
\cite{Boulware:1985wk,Myers:1987qx}. 
This displays the fact that general $S_{\cal L}$ actions contain only two derivatives, 
further reducing to just one in \sch{} coordinates.

The four independent components of $R^{\hat \mu}{}_{\hat \nu \hat \alpha \hat \beta}$ 
can be expressed as linear combinations of the three components of $R_{\hat \mu \hat \nu}$ and 
the single independent component of the Weyl tensor:
\beqa
\label{eq: Riemann from Ricci}
\label{eq: Riemann from Ricci: explicitly}
\left(\matrix{A \cr B \cr C \cr \psi}\right)
=
{1\over n(n+1)}
\left(
\matrix{
- n (F+G) + n H + n(n-1) \chi
\cr 
n F - G + H + (n-1) \chi
\cr 
- F + n G + H + (n-1) \chi
\cr 
- (F + G) + (n+2) H - 2 \chi
}\right)
.
\eeqa
From (\ref{eq: Riemann from Ricci}) 
it immediately follows that an arbitrary Riemann${}^k$ invariant 
can be generally written schematically as
\beqa
\label{eq: general Riemann^k from Ricci}
(R_{\hat * \hat * \hat * \hat *})^k = 
a \chi^k + \chi^{k-1} (R_{\hat * \hat *}) + \chi^{k-2} (R_{\hat * \hat *})^2 + ... + (R_{\hat * \hat *})^k
.
\eeqa
In this fashion, powers of $\chi$ multiplying structures involving only 
components of the Ricci tensor $R_{\hat * \hat *}$
keep track of the number of Weyl tensors used in constructing the $(R_{\hat * \hat * \hat * \hat *})^k$.
In particular, $\LL_k$ 
can be described in an alternate way to (\ref{eq: lovelock: evaluated}), 
\beqa
\label{eq: Lovelock from Ricci}
\LL_k = 
a_k^n \chi^k + \chi^{k-1} (R_{\hat * \hat *}) + \chi^{k-2} (R_{\hat * \hat *})^2 + ... + (R_{\hat * \hat *})^k
\eeqa
with 
$
a_k^n =
{(n+1)! (k-1) (n k + k - n - 2)
\over (2 k)! (n-2k+2)!}
\left({- 4 \over n(n+1) }\right)^k
$, 
and, as an immediate corollary of 
(\ref{eq: general Riemann^k from Ricci}) and (\ref{eq: Lovelock from Ricci})
\beqa
\label{eq: general Riemann^k from Lovelock}
(R_{\hat * \hat * \hat * \hat *})^k &=& 
{a \over a_k^n} \LL_k + \chi^{k-1} (R_{\hat * \hat *}) + \chi^{k-2} (R_{\hat * \hat *})^2 + ... + (R_{\hat * \hat *})^k
.
\eeqa
Any Riemann${}^k$ invariant is proportional to $\LL_k$ modulo
terms involving at least one occurrence of $R_{\hat * \hat *}$. 
In general, all powers of $\chi$ will be present, 
so it is not possible to express arbitrary scalars in terms of only Riccis
and the highest Lovelock.

\section{Specific examples and general patterns}
\label{section: examples}

Consider now some explicit examples.
The most trivial one is $k=0$, the constant (cosmological) term
\beqa
\label{eq: L0}
\LL_0 = 1
.
\eeqa
%
At $k=1$, the Ricci scalar is the unique invariant 
\beqa
\label{eq: L1}
\LL_1 = R
.
\eeqa

At $k=2$, there are three invariants:
\beqa
\label{eq: R2}
I^2_1 \equiv R^2
,~
I^2_2 \equiv R_{\mu\nu} R^{\mu\nu}
,~
I^2_3 \equiv R_{\mu\nu\alpha\beta} R^{\mu\nu\alpha\beta}
,~
\eeqa
whose Lovelock combination (\ref{eq: general lovelock}) is 
\beqa
\label{eq: L2}
\LL_2 = 
{1\over 6} 
\left[
R_{\mu\nu\alpha\beta} R^{\mu\nu\alpha\beta} - 4 R_{\mu\nu} R^{\mu\nu} + R^2
\right]
.
\eeqa
There are still 3 invariants we can construct out of Ricci and Weyl:
\beqa
\label{eq: riemann squared}
\tr R^2 = F^2 + G^2 + n H^2
,\quad
(\tr R)^2 = (F + G + n H)^2
;\quad
\tr C^2 = 4 \chi^2 {n-1 \over n+1} 
.
\eeqa

At order $k=3$, the basis of curvature invariants 
(for general metrics) 
has eight members \cite{Fulling92}:
\beqa
\label{eq: R3}
&&
I^3_1 \equiv R^3 
,\quad
I^3_2 \equiv R R_{\mu\nu} R^{\mu\nu} 
,\quad
I^3_3 \equiv R_{\nu\alpha} R^\nu{}_\mu R^{\alpha\mu} 
,\nonumber\\&&
I^3_4 \equiv R_{\nu\alpha} R_{\mu\beta} R^{\nu\mu\alpha\beta} 
,\quad
I^3_5 \equiv R R_{\mu\nu\alpha\beta} R^{\mu\nu\alpha\beta} 
,\quad
I^3_6 \equiv R_{\nu\alpha} R_{\beta\gamma\e}{}^\nu R^{\beta\gamma\e\alpha} 
,\nonumber\\&&
I^3_7 \equiv R_{\mu\nu\alpha\beta} R^{\mu\nu}{}_{\gamma\e} R^{\alpha\beta\gamma\e}
,\quad
I^3_8 \equiv R_{\mu\nu\alpha\beta} R^\mu{}_\gamma{}^\alpha{}_\e R^{\nu\gamma\beta\e} 
.
\eeqa
Only $I^3_7$ and $I^3_8$ contain $C^3$. 
One linear combination of these can be completed to 
\beqa
\label{eq: L3}
\LL_3 &=& 
{1\over 90}
\left[
I^3_1 - 24 I^3_2 + 16 I^3_3 + 24 I^3_4 + 3 I^3_5 - 24 I^3_6 + 4 I^3_7 - 8 I^3_8
\right]
.
\eeqa
The other combination of $I^3_7$ and $I^3_8$, 
\beqa
\label{eq: O3}
\OO_3 &=& 
4 I^3_1 - 12 (n+1) I^3_2 + 8 (3n-1) I^3_3 
\nonumber\\&&
+ 12 (n^2-n+2) I^3_4 + 6 (n-1) I^3_5 - 12 (n^2-1) I^3_6 
\nonumber\\&&
- 4 (n^3 - 3 n^2 + 2 n + 2) I^3_7 + (3n^2 - 3n - 2) I^3_8
\eeqa
can be shown to vanish for our geometries 
as required by the uniqueness arguments of the previous section,
see (\ref{eq: general Riemann^k from Lovelock}).
The remaining seven independent invariants 
have the explicit form:
\beqa
\label{eq: riemann cube}
\tr R^3 &=& F^3 + G^3 + n H^3
,\nonumber\\
(\tr R) (\tr R^2) &=& (F + G + n H) (F^2 + G^2 + n H^2)
,\nonumber\\
(\tr R)^3 &=& (F + G + n H)^3
; \nonumber\\
C^{\mu\nu}{}_{\alpha\beta} R_\mu{}^\alpha R_\nu{}^\beta &=&
- 2 \chi {n-1 \over n+1} (F - H) (G - H)
; \nonumber\\
(C^2)^{\mu\nu}{}_{\alpha\nu} R_\mu{}^\alpha &=& 
2 \chi^2 {n-1 \over n(n+1)} \left[ (F + G)(n-1) +2 H \right]
,\nonumber\\
(\tr C^2) (\tr R) &=& 4 \chi^2 {n-1 \over n+1} (F + G + n H)
; \nonumber\\
(\tr C^3) &=& 
\left( -2 \chi {n-1 \over n+1} \right)^3
\left[ 1 - {2\over n^2} + {4\over n^2(n-1)^2} 
\right]
.
\eeqa

At order $k=4$, there are in general 26 invariants \cite{Fulling92}, which we do not list here.
Of these, seven contain $C^4$, 
the single surviving combination of which can be completed 
\cite{Myers:1987qx,Wheeler}
to make
\beqa
\label{eq: L4}
\LL_4 &=& 
96
\left[
\half (R^{\mu\nu}{}_{\alpha\beta} R^{\alpha\beta}{}_{\mu\nu})^2 
+ R^{\mu\nu}{}_{\alpha\beta} R^{\alpha\beta}{}_{\e\gamma} R^{\e\gamma}{}_{\kappa\zeta} R^{\kappa\zeta}{}_{\mu\nu}  
+ 8 R^{\mu\nu}{}_{\alpha\beta} R^{\beta\gamma}{}_{\nu\e} R^{\e\kappa}{}_{\gamma\zeta} R^{\zeta\alpha}{}_{\kappa\mu}  
\right.\nonumber\\&&\quad\left.
- 8 R^{\mu\nu}{}_{\e\gamma} R^{\gamma\kappa}{}_{\mu\nu} R^{\alpha\beta}{}_{\kappa\zeta} R^{\zeta\e}{}_{\alpha\beta}  
+ 16 R^{\e\kappa}{}_{\mu\nu} R^{\mu\nu}{}_{\alpha\beta} R^{\alpha\zeta}{}_{\e\gamma} R^{\beta\gamma}{}_{\kappa\zeta}  
\right.\nonumber\\&&\quad\left.
+ 16 R^{\alpha\kappa}{}_{\mu\nu} R^{\mu\e}{}_{\alpha\beta} R^{\nu\zeta}{}_{\e\gamma} R^{\beta\gamma}{}_{\kappa\zeta}  
+ (\mbox{terms involving $R$ and $R_\mu{}^\nu$})  
\right]
.
\eeqa
Of the remaining 19 terms, 
six more also vanish.
Finally, we are left with 14 linearly independent invariants, including all possible non-Weyl terms:
\beqa
\label{eq: riemann fourth}
&&
\tr R^4 
,\quad 
(\tr R^3) (\tr R)
,\quad 
(\tr R^2)^2
,\quad 
(\tr R^2) (\tr R)^2
,\quad 
(\tr R)^4
; \nonumber\\&&
C^{\mu\nu}{}_{\alpha\beta} (R^2)_\mu{}^\alpha R_\nu{}^\beta
,\quad 
C^{\mu\nu}{}_{\alpha\beta} R_\mu{}^\alpha R_\nu{}^\beta (\tr R)
; \nonumber\\&&
(\tr C^2) (\tr R^2) 
,\quad 
(\tr C^2) (\tr R)^2
,\quad 
(C^2)^{\mu\nu}{}_{\alpha\nu} (R^2)_\mu{}^\alpha 
,\quad 
(C^2)^{\mu\nu}{}_{\alpha\beta} R_\mu{}^\alpha R_\nu{}^\beta
;\nonumber\\&&
(C^3)^{\mu\nu}{}_{\alpha\nu} R_\mu{}^\alpha 
,\quad 
(\tr C^3) (\tr R) 
; \nonumber\\&&
\tr C^4
.
\eeqa


This is the general pattern. 
Maximal reduction occurs for Weyl${}^k$;
any possible contraction of indices in the $k$ occurrences of the Weyl tensor
is proportional to a single invariant 
\beqa
\label{eq: weyl only invariants}
(\tr C^k) \propto \chi^k
.
\eeqa
With one Ricci, there are two independent invariants
\beqa
\label{eq: weyl and one ricci invariants}
(C^k)^{\mu\nu}{}_{\alpha\nu} R_\mu{}^\alpha 
,\quad 
(\tr C^k) (\tr R) 
;
\eeqa
there are no others since the only 
combinations linear in $F$, $G$, $H$ and symmetric under $F \lra G$ are
$(F+G)$ and $H$.
Similarly, there can be at most four invariants with two Riccis, since 
the $F \lra G$ symmetric bilinears in $F, G, H$ are 
$F^2 + G^2$, 
$F G$, 
$(F + G) H$, and
$H^2$;
for example one can take
\beqa
\label{eq: weyl and two ricci invariants}
(\tr C^{k-2}) (\tr R^2) 
,\quad 
(\tr C^{k-2}) (\tr R)^2
,\quad 
(C^{k-2})^{\mu\nu}{}_{\alpha\nu} (R^2)_\mu{}^\alpha 
,\quad 
(C^{k-2})^{\mu\nu}{}_{\alpha\beta} R_\mu{}^\alpha R_\nu{}^\beta
.
\eeqa
For low $D$ we may get fewer linearly independent invariants.
Thus, in the case of $D=4$ the identity 
$(C^2)^{\mu\alpha}{}_{\nu\alpha} = 
{1 \over 4} \delta_\nu^\mu (C^2)^{\alpha\beta}{}_{\alpha\beta}$
relates the two scalars in (\ref{eq: weyl and one ricci invariants}),
and there are similar relations in (\ref{eq: weyl and two ricci invariants}).
Indeed, $D=4$ is both special, and of course the most extensively studied 
\cite{Kramer:1980zn}.
The greatest variety occurs for scalars made entirely out of Ricci, 
which are of the form
\beqa
\label{eq: ricci only invariants}
(\tr R^{k_1}) (\tr R^{k_2}) (\tr R^{k_3}) ...
,\quad 
k_1 + k_2 + k_3 + ... = k
.
\eeqa
Here, $k_i \le D$ 
since the characteristic polynomial provides a relation, 
$0 = c_R(R)= R^D + c_1 R^{D-1} + c_2 R^{D-2} + ... + c_{D-1} R + c_D$
which expresses $R^D$ in terms of lower powers of $R$.
Moreover, since $n=D-2$ out of $D$ eigenvalues of $R$ are degenerate, 
vanishing discriminants provide additional relations. 
For example, the discriminant of $c_R(R)$, namely 
${\cal D} \equiv \prod_{i < j} (\lambda_i - \lambda_j)^2
= {\cal D} (c_1, ... , c_D)$ 
where 
$\lambda_i$ are the eigenvalues of $R$, is a homogeneous polynomial 
in the entries of $R$, of degree $D(D-1)$.
When some of the $\lambda_i$'s coincide, ${\cal D} (c_1, ... , c_D)=0$ and 
we have an additional relation on the order $k=D(D-1)$ invariants.
In fact, since $n=D-2$ eigenvalues coincide, 
there are $n(n-1)$ such relations for $k = D(D-1), D(D-1)-2, D(D-1)-4$, etc.

We see that things get more complicated
with growing powers of Ricci. 
A crude upper bound on the number of possible invariants 
made with $l$ Riccis (\ref{eq: Ricci: projectors}) 
and $(k-l)$ Weyls (\ref{eq: Weyl: projectors}), 
is the number of linearly independent 
homogeneous polynomials of order $l$ in the three variables $F$, $G$, and $H$
which are symmetric under $F \lra G$.

\section{Less symmetry?}
\label{section: cylindrical}

So far we have relied heavily on maximal symmetry.
In this section we show why this is so special by 
considering the next simplest case of 
static axially symmetric $D=n+4$-dimensional metrics, 
\beqa
\label{eq: cylindrical ansatz}
ds^2 &=& -e^{2U} (dt + {\cal A} d\phi)^2 + e^{-2 U} r^2 d\phi^2 + e^{-2 V} (dr^2 + dz^2) + e^{-2 Y} d\Omega_n^2
\nonumber\\
&=& - \omega_{\hat t}^2 + \omega_{\hat \phi}^2 + \omega_{\hat r}^2 + \omega_{\hat z}^2 + \sum_{i=1}^n \omega_{\hat i}^2
;
\eeqa
the four functions ${\cal A}$, $U$, $V$, and $Y$ depend on both $r$ and $z$. 
The problem now becomes much less tractable.
The connection one-forms (\ref{eq: curvature 2 form}) are
\beqa
\label{eq: cylindrical curvature forms}
&&
\omega_{\hat t \hat r} = e^V U_{,r} \omega_{\hat t} + {1\over 2 r} e^{V+2U} {\cal A}_{,r} \omega_{\hat \phi}
,~
\omega_{\hat t \hat z} = e^V U_{,z} \omega_{\hat t} + {1\over 2 r} e^{V+2U} {\cal A}_{,z} \omega_{\hat \phi}
,~
\nonumber\\&&
\omega_{\hat t \hat \phi} = -{1\over 2 r} e^{V+2U} ({\cal A}_{,r} \omega_{\hat r} + {\cal A}_{,z} \omega_{\hat z})
,~
\omega_{\hat r \hat z} = e^V V_{,z} \omega_{\hat r} + e^V V_{,r} \omega_{\hat z}
,~
\nonumber\\&&
\omega_{\hat r \hat \phi} = {1\over 2 r} e^{V+2U} {\cal A}_{,r} \omega_{\hat t} + e^V (U_{,r} - {1\over r}) \omega_{\hat \phi}
,~
\omega_{\hat z \hat \phi} = {1\over 2 r} e^{V+2U} {\cal A}_{,z} \omega_{\hat t} + e^V U_{,z} \omega_{\hat \phi}
,~
\nonumber\\&&
\omega_{\hat r \hat i} = e^V Y_{,r} \omega_{\hat i}
,~
\omega_{\hat z \hat i} = e^V Y_{,z} \omega_{\hat i}
,~
\omega_{\hat t \hat i} = \omega_{\hat \phi \hat i} = 0
.
\eeqa
It has the following block diagonal form:
\beqa
\label{eq: Ricci: cylindrical}
R_{\hat \mu \hat \nu} =
\left( \matrix{
R_{\hat t \hat t} & R_{\hat t \hat \phi} & & & \cr
R_{\hat \phi \hat t} & R_{\hat \phi \hat \phi} & & & \cr
& & R_{\hat r \hat r} & R_{\hat r \hat z} & \cr
& & R_{\hat z \hat r} & R_{\hat z \hat z} & \cr
& & & & \psi_n \bone_{n \times n} 
}\right)
\eeqa
with all other entries zero.
The Ricci tensor has 7 independent components, at least for $D \ge 5$:
\beqa 
R_{\hat t \hat t} &=& 
e^{2 V} 
\left[
{e^{4 U} \over 2 r^2}
({\cal A}_{,r}^2 + {\cal A}_{,z}^2)
+ {1\over r} U_{,r}
+ (U_{,rr} + U_{,zz})
- n (U_{,r} Y_{,r} + U_{,z} Y_{,z})
\right],
\nonumber\\
R_{\hat \phi \hat \phi} &=& 
e^{2 V} 
\left[
{e^{4 U} \over 2 r^2}
({\cal A}_{,r}^2 + {\cal A}_{,z}^2)
+ {1\over r} U_{,r} 
+ (U_{,rr} + U_{,zz})
- n (U_{,r} Y_{,r} + U_{,z} Y_{,z} - {1\over r} Y_{,r} )
\right]
,\nonumber\\
R_{\hat t \hat \phi} = R_{\hat \phi \hat t} &=& 
{e^{2 (U+V)} \over 2 r}
\left[
({\cal A}_{,rr} + {\cal A}_{,zz})
- {1\over r} {\cal A}_{,r}
+ 4 (U_{,r} {\cal A}_{,r} + U_{,z} {\cal A}_{,z})
- n (Y_{,r} {\cal A}_{,r} + Y_{,z} {\cal A}_{,z})
\right]
;
\!\!
\nonumber\\
R_{\hat r \hat r} &=& 
e^{2 V} 
\left[
{e^{4 U} \over 2 r^2} {\cal A}_{,r}^2 
+ (V_{,rr} + V_{,zz}) - 2 U_{,r}^2 
+ {1\over r} ( 2 U_{,r} - V_{,r}) 
\right.\nonumber\\&&\hspace{7em}\left.
+ n (Y_{,rr} - Y_{,r}^2 + V_{,r} Y_{,r} - V_{,z} Y_{,z})
\right]
,\nonumber\\
R_{\hat z \hat z} &=& 
e^{2 V} 
\left[
{e^{4 U} \over 2 r^2} {\cal A}_{,z}^2 
+ (V_{,rr} + V_{,zz}) - 2 U_{,z}^2 
+ {1\over r} V_{,r} 
\right.\nonumber\\&&\hspace{7em}\left.
+ n (Y_{,zz} - Y_{,z}^2 + V_{,z} Y_{,z} - V_{,r} Y_{,r})
\right]
,\nonumber\\
R_{\hat r \hat z} = R_{\hat z \hat r} &=& 
e^{2 V} 
\left[
{e^{4 U} \over 2 r^2} {\cal A}_{,r} {\cal A}_{,z}
+ {1\over r} ( U_{,z} - V_{,z}) - 2 U_{,r} U_{,z} 
\right.\nonumber\\&&\hspace{7em}\left.
+ n (V_{,r} Y_{,z} + V_{,z} Y_{,r} - Y_{,r} Y_{,z} + Y_{,rz})
\right]
; 
\nonumber\\
\psi_n &=&
(n-1) e^{2 Y} + 
e^{2 V}
\left[
{1\over r} Y_{,r} + (Y_{,rr} + Y_{,zz}) - n (Y_{,r}^2 + Y_{,z}^2)
\right]
.
\eeqa
More importantly, the Weyl tensor no longer depends on a single function,
so one can construct more than one
linearly independent combination of Weyl${}^k$ up to terms involving Riccis, 
and there is no longer any decomposition of the form 
(\ref{eq: general Riemann^k from Lovelock}).
So indeed spherical symmetry is very crucial for our simplifications.

\section{Effective actions}
\label{section: discussion}

For the purpose of effective actions,
there are basically three types of terms at any order $k$. 
The first are the pure Lovelock invariants $\LL_k$, with actions (\ref{eq: nice action}); 
their solutions \cite{Myers:1987qx} are direct
extensions of the single Gauss-Bonnet metric \cite{Boulware:1985wk}. 
At the other extreme, all actions with two or more Ricci tensors 
permit Ricci-flat, i.e. \sch{} solutions, 
since the Euler-Lagrange equations always contain at least one Ricci tensor or scalar.
Indeed, they may even allow \sch-deSitter metrics where 
$R_{\mu\nu} = \lambda g_{\mu \nu}$ 
provided the relative coefficients $a_k$ in (\ref{eq: nice action}) are appropriately chosen.
Being of higher derivative order, they will also have 
other, non-Einstein space, solutions. 
The third, intermediate, class consists of actions of the form 
$\int (d^D x) \sqrt{-g} R_{\mu\nu} f^{\mu\nu}$,
with $f^{\mu\nu} \equiv \alpha (C^{k-1})^{\mu\nu} + \beta g^{\mu\nu} \tr C^{k-1}$.
These actions 
do {\it not} have Ricci-flat solutions: 
$R_{\mu\nu}=0$ implies \sch, 
completely determining $C^{\mu\nu}{}_{\alpha\beta}$ as well; 
the resulting $f^{\mu\nu}$ is then incompatible with the (Ricci-flat) equations of motion 
\beqa
\n_\alpha \n^\mu f^{\alpha\nu} + \n_\alpha \n^\mu f^{\nu\alpha} 
- \n_\alpha \n^\alpha f^{\mu\nu} - g^{\mu\nu} \n_\alpha \n_\beta f^{\alpha\beta} = 0
.
\eeqa
We have not been able to find explicit solutions here, 
apart from the obvious ($C^{\mu\nu}{}_{\alpha\beta}=0$) vacua: arbitrary (A)dS spaces.

To summarize, we have presented a framework 
for categorizing the local invariants of 
static spherically symmetric geometries. 
This made possible, in particular, 
a simple division, into three general classes, 
of the local actions beyond Einstein's,
such as those implied by string theory expansions. 
These were distinguished by their Ricci tensor dependence. 
It would be interesting to see if any of these candidates 
give rise to novel, post-\sch/Lovelock effects.

\section{Acknowledgements}

We thank Marta Gomez-Reino for computer help.
This work was supported in part by
NSF grant PHY04-01667.

\setcounter{equation}{0}
\setcounter{footnote}{0}
\setcounter{section}{0}

\appendix{Powers of Curvature}
\label{section: invariants: explicitly}

In this Appendix we list some invariants used in 
Sections \ref{section: projectors} and \ref{section: general lovelock}.
Powers of Ricci, Riemann and Weyl are given by
\beqa
\label{eq: powers of Ricci: projectors}
(R^k)_\alpha{}^\beta 
&\equiv& 
R_\alpha{}^{\alpha_2} R_{\alpha_2}{}^{\alpha_3} ... R_{\alpha_k}{}^\beta 
= 
(F^k \tau + G^k \rho + H^k \sigma )_\alpha^\beta
,
\\
\label{eq: powers of Riemann: projectors}
(R^k)^{\alpha\beta}{}_{\mu\nu} &\equiv& R^{\alpha\beta}{}_{\e_2 \gamma_2} 
R^{\e_2 \gamma_2}{}_{\e_3 \gamma_3} ... R^{\e_k \gamma_k}{}_{\mu\nu} 
\nonumber
\\
&=& 2^k
\left[ 
(-A)^k \cdot 2 \tau_{[\mu}^{[\alpha} \rho_{\nu]}^{\beta]} 
+ B^k \cdot 2 \tau_{[\mu}^{[\alpha} \sigma_{\nu]}^{\beta]} 
+ C^k \cdot 2 \rho_{[\mu}^{[\alpha} \sigma_{\nu]}^{\beta]} 
+ \psi^k \cdot \sigma_{[\mu}^{[\alpha} \sigma_{\nu]}^{\beta]}
\right]
,
\!\!\!
\\
\label{eq: powers of Weyl: projectors}
(C^k)^{\alpha\beta}{}_{\mu\nu} &\equiv& 
C^{\alpha\beta}{}_{\e_2 \gamma_2} C^{\e_2 \gamma_2}{}_{\e_3 \gamma_3} ... C^{\e_k \gamma_k}{}_{\mu\nu} 
\\
\nonumber
&=& \left( -2 \chi {n-1 \over n+1} \right)^k
\left[ 
2 \tau_{[\mu}^{[\alpha} \rho_{\nu]}^{\beta]} 
+ 2 \left( - {1\over n} \right)^k \left( \tau_{[\mu}^{[\alpha} \sigma_{\nu]}^{\beta]} + \rho_{[\mu}^{[\alpha} \sigma_{\nu]}^{\beta]} \right)
+ \left( {2\over n(n-1)} \right)^k \sigma_{[\mu}^{[\alpha} \sigma_{\nu]}^{\beta]}
\right]
.
\hspace{-1em}
\eeqa
Here, $k=0$ is the identity tensor of the appropriate symmetry type, e.g. 
$(R^0)_\alpha{}^\beta = \delta_\alpha{}^\beta$ 
and
$(R^0)^{\alpha\beta}{}_{\mu\nu} = 
2 \tau_{[\mu}^{[\alpha} \rho_{\nu]}^{\beta]} 
+ 2 \tau_{[\mu}^{[\alpha} \sigma_{\nu]}^{\beta]} 
+ 2 \rho_{[\mu}^{[\alpha} \sigma_{\nu]}^{\beta]} 
+ \sigma_{[\mu}^{[\alpha} \sigma_{\nu]}^{\beta]}
$. 

Contraction of these produces 
\beqa
\label{eq: powers of Ricci: fully contracted}
(R^k)_\alpha{}^\alpha 
&=& 
F^k + G^k + H^k n
,
\\
\label{eq: powers of Riemann: contracted}
(R^k)^{\alpha\beta}{}_{\gamma\beta} 
&=& 2^{k-1}
\left[ 
  \tau   \left( (-A)^k + n B^k \right) 
+ \rho   \left( (-A)^k + n C^k \right) 
+ \sigma \left( B^k + C^k + (n-1) \psi^k \right)
\right]_{\gamma}^{\alpha}
\nonumber\\
\\
\label{eq: powers of Weyl: contracted}
(C^k)^{\alpha\beta}{}_{\gamma\beta} 
&=& \half \left( -2 \chi {n-1 \over n+1} \right)^k
\left[ 
(\tau + \rho) \left( 1 - ( - n)^{1-k} \right) 
+ {2 \sigma \over n^k} \left( ({2\over n-1})^{k-1} - (-1)^{k-1}  \right)
\right]_{\gamma}^{\alpha}
\nonumber\\
\eeqa
and in particular at $D=4$, we recover the identity
\beqa
\label{eq: Weyl squared: D=4}
(C^2)^{\alpha\beta}{}_{\gamma\beta} \Big|_{n=2}
&=& 
\left( -2 \chi {1 \over 3} \right)^2
{3 \over 4}
\left[ 
\tau + \rho + \sigma
\right]_{\gamma}^{\alpha}
=
{3 \over 4} \delta_{\gamma}^{\alpha} 
\left( -2 \chi {1 \over 3} \right)^2
=
{1 \over 4} \delta_{\gamma}^{\alpha} 
(C^2)^{\mu\nu}{}_{\mu\nu} 
.
\eeqa

Some examples of resulting invariants are
\beqa
\label{eq: Weyl^k}
(C^k)^{\alpha\beta}{}_{\alpha\beta} 
&=& \left( -2 \chi {n-1 \over n+1} \right)^k
\left[ 
1 - {2 (-1)^{k-1} \over n^{k-1}} 
+ {2^{k-1}\over n^{k-1}(n-1)^{k-1}} 
\right]
,\\
\label{eq: Weyl^k Ricciscalar}
(C^k)^{\alpha\beta}{}_{\alpha\beta} R^\mu{}_\mu
&=& \left( -2 \chi {n-1 \over n+1} \right)^k
\left[ 
1-{2 (-1)^{k-1} \over n^{k-1}} 
+ {2^{k-1}\over n^{k-1}(n-1)^{k-1}} 
\right]
(F+G+nH)
,\\
\label{eq: Weyl^k Ricci other}
(C^k)^{\alpha\beta}{}_{\gamma\beta} R^\gamma{}_\alpha 
&=& \left( -2 \chi {n-1 \over n+1} \right)^k
\left[ 
{F+G \over 2}\left(1-{(-1)^{k-1} \over n^{k-1}} \right)
+ {H \over n^{k-1}} 
\left( {2^{k-1}\over (n-1)^{k-1}} - (-1)^{k-1} \right)
\right]
.
\hspace{-2em}\nonumber\\
\eeqa 
To get the equivalents of (\ref{eq: Weyl^k Ricciscalar}) and (\ref{eq: Weyl^k Ricci other}) 
for $R^k$ instead of $R$, 
all we have to do is replace 
$F$, $G$, and $H$ in their right hand sides by $F^k$, $G^k$, and $H^k$ respectively.

\end{document}